\documentclass[sigconf,screen]{acmart}
\acmConference[ICSE 2024]{46th International Conference on Software Engineering}{April 2024}{Lisbon, Portugal}
\usepackage[utf8]{inputenc}

\usepackage{soul}
\usepackage{xcolor}
\usepackage{xspace}
\usepackage{url}
\usepackage{graphicx}
\usepackage{listings,multicol}
\usepackage[caption=false]{subfig}  
\usepackage[export]{adjustbox}
\usepackage{amsmath}

\usepackage{amssymb}
\usepackage{textcomp}
\usepackage{breakurl} 
\usepackage{listings}
\usepackage{hyperref}
\usepackage{cleveref}
\usepackage{algorithm}
\usepackage[noend]{algpseudocode}
\usepackage{dblfloatfix} 
\usepackage{enumitem}
\usepackage{kantlipsum}
\usepackage{multirow}

\title{Bridging Education and Development:\\IDEs as Interactive Learning Platforms}

\begin{document}

\author{Anastasiia Birillo}
\email{anastasia.birillo@jetbrains.com}
\affiliation{%
  \institution{JetBrains Research}
  \city{Belgrade}
  \country{Serbia}
}

\author{Maria Tigina}
\email{maria.tigina@jetbrains.com}
\affiliation{%
  \institution{JetBrains Research}
  \city{Belgrade}
  \country{Serbia}
}

\author{Zarina Kurbatova}
\email{zarina.kurbatova@jetbrains.com}
\affiliation{%
  \institution{JetBrains Research}
  \city{Belgrade}
  \country{Serbia}
}

\author{Anna Potriasaeva}
\email{anna.potriasaeva@jetbrains.com}
\affiliation{%
  \institution{JetBrains Research}
  \city{Belgrade}
  \country{Serbia}
}

\author{Ilya Vlasov}
\email{ilya.vlasov@jetbrains.com}
\affiliation{%
  \institution{JetBrains Research}
  \city{Yerevan}
  \country{Armenia}
}

\author{Valerii Ovchinnikov}
\email{vovchinnikov02@gmail.com}
\affiliation{%
  \institution{Constructor University}
  \city{Bremen}
  \country{Germany}
}

\author{Igor Gerasimov}
\email{igor.gerasimov@jetbrains.com}
\affiliation{%
  \institution{JetBrains}
  \city{Berlin}
  \country{Germany}
}

\begin{abstract}

In this work, we introduce a novel approach to programming education -- in-IDE courses implemented for IntelliJ-based IDEs via the JetBrains Academy Plugin.
The primary objective of this approach is to address the challenge of familiarizing students with industrial technologies by moving all theory and practical materials to a professional IDE. This approach allows students to immediately use modern industrial tools as they are fully integrated into the learning process. 
We have already applied this approach in over 40 courses, and it successfully educates students across diverse topics such as Plugin Development, Algorithms, Data Analysis, and Language mastery in various programming languages, including Kotlin, Java, C++, and Python.
Along with the paper, we are providing the community not only with a new way of learning and a set of ready-made courses but also a collection of helpful resources to assist educators in getting started with the plugin. Finally, we describe in detail an IDE plugin development course that demonstrates how the in-IDE approach covers complex topics easily.

\end{abstract}

\keywords{programming education, MOOC, in-IDE learning, IDE plugins}

\maketitle

\section{Introduction}

The ever-evolving field of computer science continually raises the demands on programming education~\cite{al2023computer}.
There is a noticeable gap between studying software engineering and real-world coding, which surfaces when transitioning from student to professional programming. Students often lack sufficient programming skills and experience using professional developer tools such as Integrated Development Environments (IDE)~\cite{radermacher2014investigating}. This leads to rigorous efforts to integrate these tools into the learning processes.

There are several approaches to integrating IDE into the learning process. One of them is embedding directly into Massive Open Online Courses (MOOCs). However, in this case, IDEs are cut in capabilities and features, reduced to a simple text editor with highlighting and basic buttons like ``Run'' or ``Compile'' the program~\cite{alario2018study}. The alternative strategy of tailoring IDEs for students is developing simplified versions or even new IDEs with fewer features, a streamlined interface, and extra guidance such as navigation tips and detailed explanations~\cite{messer2022detecting, barphe2022effective, gross2005evaluating}. However, even with these adjustments, students will still need to make some effort to get conversant with professional-grade IDEs after the educational process~\cite{radermacher2014investigating}.

To address these challenges, we present a novel solution: the in-IDE JetBrains Academy education plugin~\cite{jetbrains-academy-plugin} for IntelliJ-based IDEs~\cite{ij-based-ides}. 
The JetBrains Academy plugin was developed to assist both students and educators and bring together MOOCs and professional IDE functionality in a single platform. Its setup allows students to familiarize themselves with these tools as an integral part of their learning journey, introducing them to various IDE features to streamline task execution. Furthermore, once students have learned to program within industrial tools, they do not need to exert additional effort to transition to a new environment for completing personal projects or applying for their first job. In addition, we equip educators with a comprehensive guide for course creation, complete with a wide selection of presets and templates specific to different programming languages to make the course development process faster and easier. 

The plugin has already facilitated the launch of over 40 courses, including 20 courses authored by JetBrains. The courses cater to many students and cover a wide range of topics, from basic language courses to learning IDE features or solving algorithm tasks. 
One of the recent releases was a course focused on the IntelliJ IDE plugin development. This course equips students with comprehensive information and test assignments, guiding them in initiating their plugin development journey, utilizing the latest technologies. This course demonstrates how the JetBrains Academy plugin helps to cover not only classic cases, but also advanced technologies such as the Intellij Platform SDK~\cite{intellijsdkdocumentation}.
Our key contributions include:
\begin{itemize}
    \item A novel approach of in-IDE learning; 
    \item A set of materials for in-IDE course creation for Kotlin, Java, and Python languages provided together with real-world course examples in different topics~\cite{jba-links};
    \item The IDE Plugin Development Course, which is a great example of an advanced in-IDE course where you can learn how to write plugins and extend IntelliJ-based IDEs.
\end{itemize}

\section{Background}~\label{sec:background}

In the transition from academic studies to professional programming, an observed gap persists between education and real-world developer work. A prevalent issue faced by novice programmers is their insufficient prowess in utilizing IDEs and their inherent tools, notably debugging facilities~\cite{radermacher2014investigating}. Consequently, to bridge this gap, a growing number of educational platforms are incorporating IDEs or their adapted versions into their learning models.

One exemplary approach involves integrating web-based IDEs, such as Codeboard~\cite{codeboard}, as a principal code editor in MOOC platforms. Recent studies attest to Codeboard's efficacy in bolstering learner engagement in MOOCs~\cite{chan2017perceived, meyer2017fourteen, alario2018study, gallego2020analyzing}. Codeboard provides a dedicated IDE environment in a browser enabling students to write, compile, and run code. The editor can be customized with different font sizes, different themes, etc. However, its limited functionality compared to full-featured IDEs can compromise preparation for real-world coding scenarios, such as using version control systems, debugging, or refactorings~\cite{snipes2015practical, amann2016study}. 

An alternative and more gentle approach to acquaint students with IDE-based development involves the use of student-friendly IDEs, like Alice~\cite{alice}, BlueJ~\cite{kolling2003bluej}, Jeliot~\cite{levy2003jeliot}, RAPTOR~\cite{carlisle2005raptor}, which offer a less intimidating doorway into the world of programming. Each of these environments is adapted to different needs and provides various modifications to improve students' experience, such as personalized hints~\cite{messer2022detecting} or used classes visualization in Java programs~\cite{weiresteejay}. 
Although some of these environments, such as BlueJ, are quite popular for Java courses~\cite{yan2009teaching}, they still raise some difficulties for students. For example, BlueJ can be deemed as an ultralight IDE for small-scale Java development, though it lacks many advanced features compared to industry-level IDEs such as NetBeans and Eclipse~\cite{yan2009teaching}. Another drawback is the lack of integration with MOOC content and platforms. These IDEs are used as an environment for solving tasks, but students still have to switch between different sources of theory, task description, and an IDE editor.

Considering conventional educational methods, several educators currently employ special educational platforms such as GitHub Classroom~\cite{githubclassroom}. This platform facilitates materials sharing, assignment templating, universal test setups, automated checks, and monitoring of student progress. However, the assessment of code quality and overall testing typically should be run manually by students in their local environment. Moreover, students aren't restricted in their choice of development environment and solve tasks using a basic text editor, so there are no guarantees that they are learning how to operate with IDEs. In line with this aspect of education, GitHub Classroom offers several integrations with IDEs that allow educators to connect an IDE to the repository with an assignment. The main limitation here is that the integration itself is mainly with a GitHub web code editor (GitHub Codespace), as integration with Visual Studio Code is no longer supported~\cite{github-vs-code}. GitHub Codespace supports a number of industry-oriented features such as code inspections or version control integration, but it simply opens a repository and does not customize it for educational purposes.

In summary, we can observe that much research is focused on teaching the use of development tools, such as IDEs, to help students overcome challenges after graduation. However, existing solutions do not cover many industrial features or have insufficient integration with MOOCs. The current situation has led us to the conclusion that attempting to integrate educational materials directly into the IDE will help address many of the aforementioned issues.

\begin{figure*}[t]
    \centering
    \includegraphics[width=\linewidth]{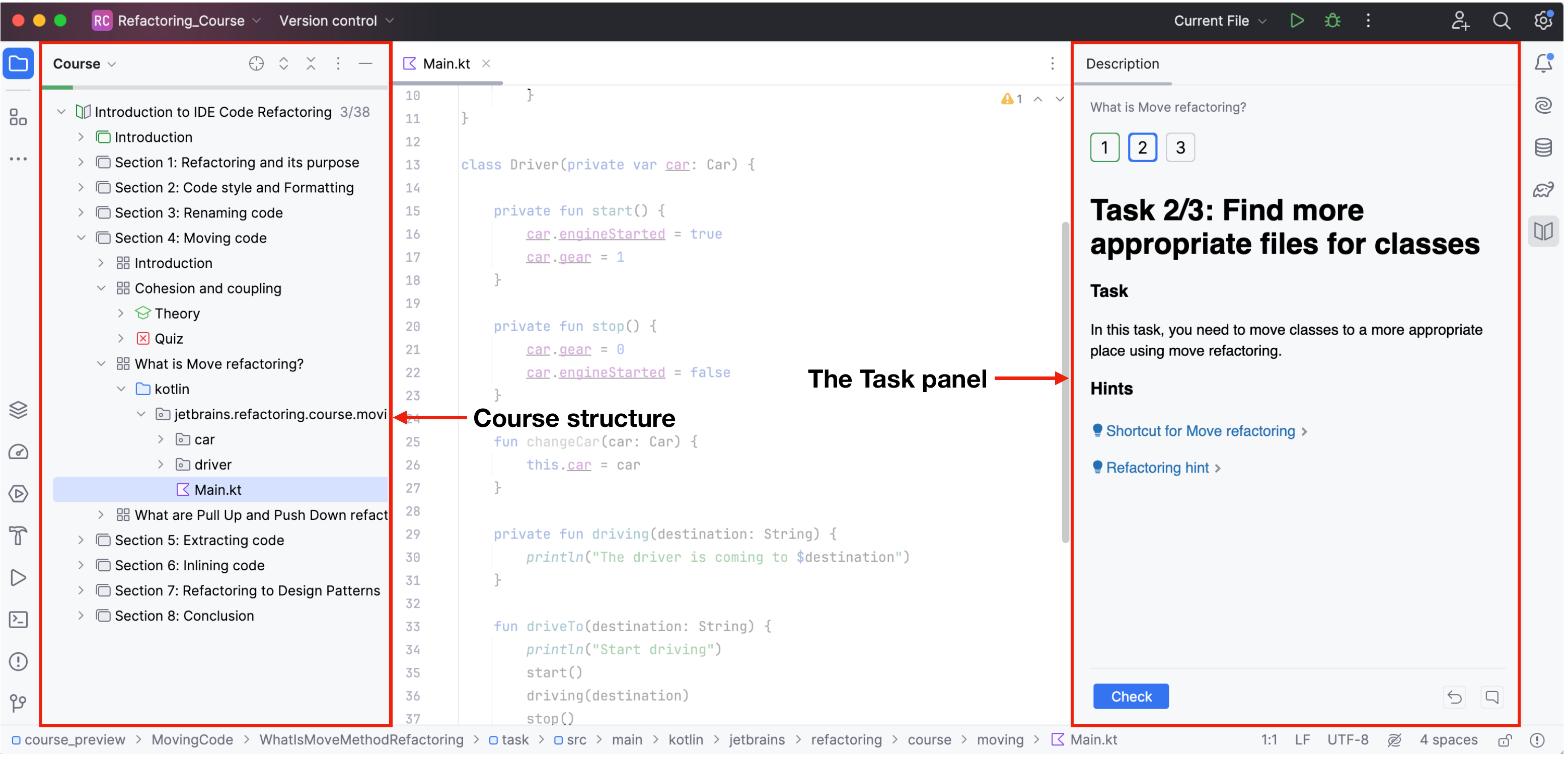}
    \vspace{-0.3cm}
    \caption{Example of JetBrains Academy plugin with the ``Introduction to IDE Code Refactoring'' course. The left side shows the course structure with the course progress itself. The right side shows the task panel with theory, task description, and hints provided by the course author.}
     \vspace{-0.3cm}
    \label{fig:examples}
\end{figure*}

\section{JetBrains Academy Plugin}~\label{sec:jba-plugin}

This section describes our approach to incorporating educational processes into the IDEs via the JetBrains Academy plugin as well as courses created with the plugin and authored by JetBrains.

\subsection{Plugin Overview}

The JetBrains Academy plugin is a plugin for IntelliJ-based IDEs such as IntelliJ IDEA, PyCharm, and CLion that allows students to learn and create interactive courses within the IDE. It is a project-based learning platform that utilizes ``learn by doing'' approach, allowing learners to explore new topics and immediately apply them in practice by solving tasks~\cite{iftikhar2022practice}.
The plugin enhances the IDE interface by displaying the Course Structure on the left side and the Task panel on the right side  (see \Cref{fig:examples}).

The course itself is represented as a project that can be downloaded and uploaded elsewhere. Every course is divided into sections that contain lessons with theoretical explanations and practical exercises. The JetBrains Academy plugin provides two types of lessons: \textit{lessons} that represent independent theory and practice, and \textit{framework lessons} that allow to work continuously on the same code through multiple tasks and exercises.
Educators can implement tests for tasks to check the learner's solution and provide customized feedback on them in both types of lessons.

\Cref{fig:examples} demonstrates an example of the framework lesson. Completed tasks are green, failed tasks are red, and not-started tasks are gray. The course structure is visualized as a course tree and helps students to track their progress. 
The task panel is the main place to interact with the course: read theory, task descriptions, hints, and check solutions. The task description is a Markdown file that is displayed in a standard way in the IDE. The learners solve the exercises and refine the project with each task they complete. While solving tasks, students can use all IDE features while learning, e.g., perform automatic refactorings, change the IDE settings, or use a debugger to find mistakes in their code. The learners can switch from the course interface to the standard project using the course tree panel.

\vspace{-0.3cm}

\subsection{Supported Tasks}

The JetBrains Academy plugin supports several types of tasks, including \textit{theory}, \textit{quiz}, and \textit{code}. The course author can combine them in any order inside one lesson.

\textbf{\textit{Theory tasks}} represent a theory part and usually do not require any coding activities. These tasks are shown in the Task panel and can contain text, figures, and links. To complete this task, students need to read the content and switch to the next task.

\textbf{\textit{Quiz tasks}},  which are also theory-based, require students to complete a quiz to complete the task. Quizzes can be designed as a single or multiple-choice, and error messages can be customized depending on the learner's chosen option.

\textbf{\textit{Code tasks}} require students to solve programming exercises. Several options can be used. In a simple case, students only need to replace a placeholder with the correct code to complete the task. In a more difficult case, students must implement a function or a class in their own way. These tasks are tested by input-output examples or provided test cases by the course author and can be customized with different approaches and frameworks. 

\subsection{Custom tests for solutions}

For \textit{code tasks}, course authors can create custom tests to check learners' solutions. The tests can use any available test systems, e.g., \textit{unittest} framework~\cite{unittest} for Python or \textit{JUnit5} for Java or Kotlin.

As a part of this work, we developed a testing framework~\cite{kotlin-test-framework} for Java and Kotlin courses that allows analyzing code structure using the Java Reflection API~\cite{reflectionApi} under the hood. 
This test method provides educators with a powerful mechanism to hide the actual solution from the learner in the tests and to manipulate entities that should be implemented by the learner from scratch.
In addition, one of this framework's modules uses the IntelliJ API under the hood and allows performing static analysis during testing, for example, checking the formatting of a code fragment or determining whether refactoring has been performed.
The framework is already used by several courses~\cite{kotlin-onboarding-course, kotlin-onboarding-oop-course, ide-code-refactoring-course}.

\vspace{-0.3cm}

\subsection{Customizing IDE}

As the course is implemented in the form of a project, the educators can customize various aspects of the IDE to help learners in the educational process.

\textbf{\textit{Code style}}. The educators can change the IDE's default 
 code style rules, such as indent size, number of blank lines, or import sorting.

\textbf{\textit{Inspections}}. The educators can change IDE's inspections highlighting or severity and turn off unnecessary ones to match the level of difficulty of the course to the learner's ability.

\textbf{\textit{Run configurations}}. Educators can define multiple run configurations for the course to help learners focus on learning new things rather than figuring out how to launch a project. 

Those customizations are available for all languages and can be made using the Settings page in IDE's UI. Next, the customized configuration files are automatically created and stored in the course root. Finally, when learners install the course, these settings are automatically applied.

\subsection{How to start}

To create a course with the JetBrains Academy plugin, refer to the course creator start guide~\cite{educator-start-guide} or an introductory video tutorial~\cite{educator-start-guide-youtube}. 
Another source of inspiration and useful materials is the set of repositories of already developed courses.~\cite{jba-github} These are all public and can be used as a starting point.
Finally, to simplify the process, we also provide a collection of templates in the form of public GitHub repositories, such as Java~\cite{java-course-template}, Kotlin~\cite{kotlin-course-template}, and Python~\cite{python-course-template} templates, to expedite and streamline course development.

These templates include preconfigured projects with basic dependencies, sample code, ready-made tests, and useful links. Because these projects are set up as course projects, they not only help educators set up a new course but also give them a quick overview of how different tasks work, which features can be used, and how to simplify the course creation process. 

As a basic example, we have extended the Kotlin template with a configuration file that adapts Kotlin inspections. We built this file based on the most popular inspections in two Kotlin courses~\cite{algorithmic-challenges-in-kotlin-course, kotlin-onboarding-course}, but it can be easily modified for the educator's needs.

\subsection{Course Distribution}

The course authors have several options to distribute the course: publishing directly to GitHub, distributing it as an archive, or uploading the course to the JetBrains Marketplace~\cite{jetbrains-academy-marketplace}.
The first two options are suitable for local use of the course but require additional effort from the student to install the course. In such cases, students are required to download the course archive into the IDE themselves to initiate the learning process. We recommend using these options to test courses.
The third option, the Marketplace, is the most convenient way of distribution for both educators and students. The Marketplace offers benefits for educators, such as collecting students' feedback and analyzing course usage statistics.

From the students' point of view, the marketplace is the easiest way to download the course and start learning. Students simply open an IDE and select the course in a special window, and the IDE automatically takes care of everything else.

\subsection{Courses Overview}

The JetBrains Academy plugin provides an interactive way to learn new concepts right inside IDEs. To present an overview of existing educational courses authored by JetBrains, we categorized them into several groups based on the covered topics.

\textbf{\textit{Programming language courses}} cover main language topics and include tasks for practical application. Marketplace currently offers courses for Kotlin, Python, Java, Rust, JavaScript, Scala, and C++ languages. For example, the ``Kotlin Onboarding: Introduction''~\cite{kotlin-onboarding-course} and ``Kotlin Onboarding: Object-Oriented programming''~\cite{kotlin-onboarding-oop-course} courses are a part of a course series to introduce learners to the foundational concepts of the Kotlin language. Learners are expected to read the short theoretical part of the lesson and then apply new knowledge right in the IDE by solving small tasks. As a result of each lesson, learners develop several projects that can be used for their portfolios in the future.

\textbf{\textit{General software engineering courses}} cover important concepts such as algorithms, or network programming. These courses are more advanced and require knowledge of some programming language to solve the tasks. For example, the ``Algorithmic Challenges in Kotlin''~\cite{algorithmic-challenges-in-kotlin-course} course introduces learners to the main algorithmic techniques and provides coding challenges to practice.

\textbf{\textit{IDE features courses}} assist learners in becoming proficient with the IDEs features. For example, the ``Introduction to IDE Code Refactoring''~\cite{ide-code-refactoring-course} course introduces learners to refactorings and the IDE features that help to perform them effectively.
The course provides theory on refactoring concepts in general, demonstrates IDE automatic refactoring features, and includes tasks to practice.

\textbf{\textit{University Courses}} are used in universities as main or supplementary material to help students become familiar with the topic directly in IDE. For example, the ``Gateway to Pandas''~\cite{ide-pandas} course allows learners to study the popular Python data analysis library called Pandas. It consists of two parts. The main objective of the first section is to provide exercises on the basics of the library. The second section consists of two real-world projects in this area. This course is designed as a practical seminar assignment for a machine learning course at the Neapolis University Paphos. 

\textbf{\textit{IDE development courses}} introduces learners to the basics of creating IDE plugins, such as adding custom language support, extension points, building User Interface components, and plugin testing. For example, the ``IDE Plugin Development'' course provides a comprehensive overview of developing plugins for IntelliJ IDEA. It offers the theory and practice tasks dedicated to creating the learner's own IntelliJ-based IDE plugins. More detailed information can be found in  \Cref{sec-ide-dev-course}.

The educational courses described above cover different topics, from learning programming languages to using IDEs better and making IDE plugins. It allows learners to get new knowledge and practice right inside their IDE. All educational courses from this section are available at JetBrains Academy Marketplace~\cite{jetbrains-academy-marketplace}.

\vspace{-0.2cm}

\section{IDE Plugin Development Course}~\label{sec-ide-dev-course}

The field of IDE Plugin development for the IntelliJ Platform faces a challenge, as related documentation and tutorials are complicated and hardly accessible for newcomers. 
This is evidenced by the thousands of questions on the plugin developer forum~\cite{ij-sdk-forum} and the daily questions on the IDE development community slack~\cite{ij-sdk-slack}.
Our course is designed to bridge this gap by offering an integrated repository of beneficial links, practical advice, and illustrative examples to guide developers through common challenges encountered at the onset of plugin development for IDEs.
This section describes the \textit{expected} course structure and the current state of the course.

\vspace{-0.2cm}

\subsection{Course Overview}

The primary learning strategy in this course is to instruct students on effectively utilizing external resources, emphasizing the application of existing theory and documentation rather than redundant rewriting it into the course content.
To follow this strategy, we accomplish the course tasks with useful materials, like  IntelliJ Platform SDK Documentation~\cite{intellijsdkdocumentation} or links to useful libraries for plugin development.
As a result, students will become proficient in navigating through documentation, enabling them to apply this skill independently to real-world tasks after completing the course.
The course is going to cover various aspects of IDE plugin development, including \textit{general knowledge}, \textit{specific IDE extensions}, \textit{working with UI}, and \textit{infrastructure questions.}

\textbf{\textit{General Knowledge}} part will contain general knowledge about the Program Structure Interface (PSI) and rules for working with it. PSI is a special code representation in the form of a tree that allows code to be manipulated in a variety of ways inside IDE. PSI is the basis of plugin development for the IntelliJ Platform.

\textbf{\textit{Specific IDE Extensions}} will demonstrate common practices for utilizing static analysis in the plugin and includes tasks for creating your own IDE features such as inspections or refactorings.

\textbf{\textit{Working with UI}} will include working with custom actions, buttons, and graphical objects, mastering JCEF~\cite{jcef} features.

\textbf{\textit{Infrastructure Issues}} will be the final section of the course. This section covers the basics of plugin infrastructure, including creation from a template, publishing methods, logging, and testing. This module will allow students to go through the whole process of plugin development and start a new project easily.

As an additional topic, we are going to include several assignments on the integration of machine learning models within the IDE. These tasks cover innovative areas of development and expose students to state-of-the-art tools for efficient machine learning inference under the JVM platform. 
We hope, that the course will help in several aspects and will be:

\begin{itemize}
    \item The main entry point for learning plugin development;
    \item A set of useful resources for plugin development like IDE extensions or libraries;
    \item A practical instrument to onboard people inside and outside JetBrains who develop plugins.
\end{itemize}

\subsection{Current State of the Course}

The current phase of our project has successfully completed the first \textbf{\textit{General Knowledge}} section focusing on the PSI, a fundamental aspect of IntelliJ plugin development. 
Here's an overview of what we've covered in this first phase:

\textbf{Integration of Comprehensive Learning Resources}. We have carefully curated and embedded a wide range of learning materials directly. These resources are not limited to theoretical content; they extend to external references such as official documentation, facilitating a deeper understanding of the PSI framework.

\textbf{Interactive Learning}. To reinforce the learning objectives, we have incorporated interactive quizzes and practical exercises into the curriculum. These activities are designed to test the learner's understanding and application of the concepts discussed. The quizzes provide immediate feedback to help learners assess their understanding of the material in real-time, while the code assignments are validated by tests.

\textbf{Hands-On Experience with Sample Projects}. A unique feature of our course is the provision of sample projects that students can interact with directly. These projects come pre-configured with UI modifications, allowing students to see the immediate impact of their code in a practical, real-world context. This hands-on approach not only enhances learning but also gives students a taste of how their skills might be applied in a professional environment.

\textbf{Final project for In-Depth Application}. The culmination of the PSI chapter is a significant project that synthesises all the concepts learnt. This project is structured in a number of steps that guide students through a progressive, in-depth exploration of PSI. At the end of the project, students will not only have a theoretical understanding of PSI, but also a tangible, self-created artefact that demonstrates their newly acquired skills. 

We believe that this combined approach will help learners become more familiar with IntelliJ plugin development, and we will be releasing the next sections of the course soon.

\vspace{-0.4cm}

\section{Conclusion And Future Work}~\label{sec:conclusion}
In this paper, we presented our approach to integrating educational processes into IDEs through the JetBrains Academy plugin.
As a successful implementation example of our idea, we have described the courses that have already been executed using this platform, including IDE Plugins Development course.
By seamlessly embedding educational content within the IDE environment, we aim to create a cohesive and immersive learning experience. This approach not only simplifies access to dedicated IDE tools but also exemplifies the successful integration of educational materials directly into the development workflow. The JetBrains Academy plugin serves as a key enabler, demonstrating the effective fusion of theory and practice in programming education. 

As we continue to refine and extend this approach, we expect to make a huge contribution to the development of educational practice in software development. 
Future work includes several directions. The first direction is to make the interaction with the plugin easier and to connect the plugin to the Learning Management Systems (LMS), e.g., Moodle~\cite{moodle}. This direction includes authorisation and authentication processes through the LMS, opening the courses in the IDE through the system interface, as well as providing the analytics for progress tracking. This approach will help educators maintain their familiar environment and continue to use the systems they are already familiar with.

The second direction involves the integration of detailed analytics to refine educational content, providing educators with tools for better guidance and using data-driven insights for continuous improvement. The main goal of this direction is to help educators improve course content based on the solutions already submitted.

The last direction is to improve the overall effectiveness and accessibility of our educational approach for students, which includes UI improvements, integration of an AI Assistant~\cite{jb-ai-assistant} and an IDE feature trainer~\cite{jb-ide-features-trainer}. The integration with the AI assistant part includes the development of a next-step hint generation solution that combines the power of Large Language Models (LLMs) and static analysis provided by the IDE. This assistant will help students to get immediate personalized help in solving coding exercises. The integration with the IDE Feature Trainer will provide additional exercises to teach students how to use various IDE features such as debugging, inspection, refactoring, etc. We believe these changes will improve the student experience and allow them to learn in a more efficient way.

\section*{Acknowledgements}
We gratefully acknowledge the JetBrains Academy team for creating a plugin and providing resources essential for the development of educational courses within the IDEs.
We would also like to thank the authors of the Marketplace courses for their creativity, ingenuity, and meaningful contribution to the students' learning experience.

\bibliographystyle{ACM-Reference-Format}
\balance
\bibliography{main}
\end{document}